\documentclass[aps,pre,amsmath,twocolumn,amssymb,floatfixng,showpacs,longbibliography,
superscriptaddress,nofootinbib]{revtex4-2}
\usepackage{amsfonts}
\usepackage{amsmath}
\usepackage{amssymb}
\usepackage{latexsym}
\usepackage{array}
\newcolumntype{P}[1]{>{\centering\arraybackslash}p{#1}}
\newcolumntype{M}[1]{>{\centering\arraybackslash}m{#1}}
\usepackage{caption}
\usepackage[colorlinks,citecolor=red,urlcolor=blue,bookmarks=false,hypertexnames=true]{hyperref} 
\usepackage{xcolor}
\usepackage{bm}
\usepackage{relsize}
\usepackage{float}
\usepackage{subfloat}
\usepackage{dcolumn}
\usepackage{epsfig}
\usepackage{graphicx}
\usepackage{multirow}
\usepackage{makecell}
\usepackage{mathtools}
\usepackage[bbgreekl]{mathbbol}

\usepackage{footnote}

\begin{document}

\title{Long Time Tails in Quantum Brownian Motion of a charged particle in a magnetic field}
\author{Suraka Bhattacharjee}
 \affiliation{Raman Research Institute, Bangalore-560080, India} 
\author{Urbashi Satpathi}
\thanks{Present address: Department of Chemistry, Ben-Gurion University of the Negev, Beer-Sheva 84105, Israel}
 \affiliation{International Center for Theoretical Sciences, Tata Institute of Fundamental Research, Bangalore-560089, India}

 \author{Supurna Sinha}
 \affiliation{Raman Research Institute, Bangalore-560080, India}

\date{\today}

\begin{abstract}
 We analyse the long time tails of a charged quantum Brownian particle in a harmonic potential in the presence of a magnetic field using the Quantum Langevin Equation as a starting point. We analyse the long time tails in the position autocorrelation function, position-velocity correlation function and velocity autocorrelation function.
 We study these correlations for a Brownian particle coupled to the Ohmic and Drude baths, via position coordinate coupling. At finite temperatures we notice a crossover from a power-law to an exponentially decaying behaviour around the thermal time scale $\frac{\hbar}{k_BT}$. We analyse how the appearance
 of the cyclotron frequency in our study of a charged quantum Brownian particle affects the behaviour of the long
 time tails and contrast it with the case of a neutral quantum Brownian particle. 
\end{abstract}

\maketitle

\section{Introduction}

There has been considerable interest in the study 
of the quantum Brownian motion of a particle coupled to a heat bath. A characteristic notable feature in such studies is a power law decay of correlations in the long time domain giving rise to the emergence of long time tails \cite{grabertone,graberttwo,decosup}. The long time tails of the velocity autocorrelation function for classical Brownian motion have been extensively studied, which show that the velocity autocorrelation function falls off as t$^{-3/2}$ at long times \cite{Alder,Tongcang}. 

The long time tails were first observed in hard disk and hard sphere liquids by a computer experiment and later the boundary effects on the power law behaviours were also studied  \cite{Alder,Kai,Felderhof}. The long time tails were observed experimentally by photon correlation spectroscopy for spherical Brownian particles \cite{Paul}. Much later, optical tweezer experiments have confirmed the $t^{-3/2}$ behaviour of velocity autocorrelation in the bulk and also the $t^{-5/2}$ fall off near the boundaries \cite{Tongcang,Franosch,Simon,Jeney,Jannasch}. However, the study of long time tails of the correlation functions for quantum Brownian motion continues to remain as a challenging problem for the theorists in this field. In \cite{graberttwo}, the authors address the problem of quantum Brownian motion at low temperatures and with arbitrarily strong damping. They notice that for a harmonically bound particle the zero-temperature correlation functions display long-time tails. At finite temperatures a power-law decay at intermediate times is followed by an exponential decay.

In contrast to earlier work \cite{grabertone,graberttwo} in which the authors address the issue of decay of temporal correlations of a neutral quantum Brownian particle, in this paper we address the appearance of long time tails in the temporal decay of correlations in the dynamics of a charged quantum Brownian particle in a harmonic potential in the presence of a magnetic field via the quantum Langevin equation \cite{ford,Urbashione,surakaone}. Moreover, our study pertains to the particle coupled to Ohmic and Drude baths through position coordinate coupling.\\
\hspace*{0.2cm} Here we have three competing time scales : (i) the time scale set by the harmonic oscillator frequency $\omega_0$, (ii) the time scale set by the viscous damping rate $\gamma$ and (iii) the time 
scale set by the cyclotron frequency $\omega_c$. We analyse the decay of various correlation functions 
as a function of time and the emergence of long time tails in various temperature regimes and study how the 
appearance of the cyclotron frequency leads to certain quantitative and qualitative changes in the temporal decay of correlations. \\
\hspace*{0.2cm}In \cite{Urbashione} the growth of the mean-square displacement of a charged Brownian particle in a magnetic field was analysed via the Quantum Langevin Equation of a charged particle in a magnetic
field. More recently we had used the same framework for 
studying the Response function, position-velocity and velocity autocorrelation functions of such a charged Brownian particle \cite{surakaone}. 
Our focus in \cite{surakaone} was on a transition from an oscillatory to a monotonic behaviour as a result of a competition between the cyclotron frequency and the viscous damping rate.\\
\hspace*{0.2cm}In the present paper we go beyond these studies and in the spirit of \cite{graberttwo}, analyse in detail, the emergence of long time tails in this system. \\
The outline of the paper is as follows. In Sec. $II$ we 
introduce the Quantum Langevin Equation for a harmonically bound charged 
particle in a magnetic field as a starting point and 
define the various correlation functions under investigation: 
(i) the position auto-correlation function, (ii) the position-velocity correlation function and the (iii) velocity
auto-correlation function. We then study in detail the 
long time decay behaviour of the position auto-correlation function, the position-velocity correlation function 
and the velocity autocorrelation function in Sec. $III$. 
We study this in various temperature
domains for the underdamped ($\omega_c >> \gamma$), critically damped ($\omega_c \sim \gamma $) and
overdamped ($\omega_c << \gamma$) cases. 
We analytically investigate the transition from a power law to an exponential behaviour with time in the long time domain. We compare two bath models : (i) the Ohmic bath model and (ii) the Drude bath model and study the differences.
In Sec. $IV$ we summarise and discuss our results and finally end the paper with 
some concluding remarks in Sec. $V$.

\section{Quantum Langevin Equation:  Response Function and Correlation Functions}

The quantum generalized Langevin equation of a charged particle in the presence of a magnetic field \cite{Urbashitwo} and a harmonic oscillator potential \cite{graberttwo, malay} is given by 
\begin{eqnarray}
\notag m \ddot{\vec{r}}(t)&=&-\int\mu (t-t')\dot{\vec{r}}(t')dt'-m\omega_{0}^{2}\vec{r}(t)+\frac{q}{c}(\dot{\vec{r}}(t)\times\vec{B})\nonumber\\
&+&\vec{F}(t) 
\label{qle}
\end{eqnarray}

where, $ m $ is the mass of the particle, $ \mu(t) $ is the memory kernel, $\omega_{0}$ is the harmonic oscillator frequency, $ q $ is the charge, $ c $ is the speed of light, $ \vec{B} $ is the applied magnetic field and $ \vec{F}(t) $ is the random force with the following properties :
\begin{align}
\langle{F_{\alpha}(t)}\rangle=0
\end{align}

\begin{align}
\notag \frac{1}{2}\langle{\lbrace F_{\alpha}(t),F_{\beta}(0)}\rbrace\rangle =\frac{\delta_{\alpha \beta}}{2 \pi }\int_{-\infty}^\infty{{d\omega}Re[\mu(\omega)]}
\hbar\omega  \\ \hspace{0.2cm} \smaller{\times}  \coth(\frac{\hbar\omega}{2k_BT}) e^{-i\omega t} \label{noisecor}
\end{align}

Here $\alpha, \beta = x, y, z$ and $\delta_{\alpha \beta}$ is the Kronecker delta function.
Also 
$\mu(\omega)= \int_{-\infty}^{\infty}{dt \mu(t) e^{i\omega t}}$.
The Response Function 
$R_x(\omega)$ is given by \cite{surakaone}:
\begin{equation}
    R_x (\omega)=\mathrm{Re}R_x(\omega)+i\mathrm{Im}R_x(\omega)
    \end{equation}
where Im$R_x(\omega)$ and Re$R_x(\omega)$ represents the imaginary part and the real part of the response function respectively
which can be directly obtained from the Quantum Langevin Equation
displayed in Eq.(\ref{qle}) via Fourier transform as:
\begin{widetext}

\begin{equation}
\mathrm{Im}R_x(\omega)=\frac{1}{m}  \frac{\mathrm{Re}[K(\omega)]\left[\left(\left(\omega-\frac{\omega_0^2}{\omega}\right) + \mathrm{Im}[K(\omega)]\right)^{2}+\omega_{c}^{2}+\mathrm{Re}[K(\omega)]^{2} \right]}{\omega\left\lbrace  \left[\left(\left(\omega-\frac{\omega_0^2}{\omega}\right) + \mathrm{Im}[K(\omega)]\right)^{2}+\omega_{c}^{2}+\mathrm{Re}[K(\omega)]^{2}  \right]^{2}- 4\omega_c^{2}\left(\left(\omega-\frac{\omega_0^2}{\omega}\right) + \mathrm{Im}[K(\omega)]\right)^{2} \right\rbrace } \label{imresponsefunctionomega}
\end{equation}

\begin{equation}
  \mathrm{Re}R_x(\omega)=\frac{1}{\pi  m}P\int_{-\infty}^{\infty} \frac{\mathrm{Re}[K(\omega')]\left[\left(\left(\omega-\frac{\omega_0^2}{\omega}\right) + \mathrm{Im}[K(\omega)]\right)^{2}+\omega_{c}^{2}+\mathrm{Re}[K(\omega)]^{2} \right]}{\omega\left\lbrace  \left[\left(\left(\omega-\frac{\omega_0^2}{\omega}\right) + \mathrm{Im}[K(\omega)]\right)^{2}+\omega_{c}^{2}+\mathrm{Re}[K(\omega)]^{2}  \right]^{2}- 4\omega_c^{2}\left(\left(\omega-\frac{\omega_0^2}{\omega}\right) + \mathrm{Im}[K(\omega)]\right)^{2} \right\rbrace } d\omega'\label{kramers}
\end{equation}

where $K(\omega)$=$\mu(\omega)$/m and $\omega_c=\frac{eB}{mc}$ is the cyclotron frequency.

\end{widetext}

As we know, the imaginary part of this Response Function 
is connected to the spectral density (the Fourier transform 
of the position autocorrelation function) via the 
Fluctuation-Dissipation Theorem \cite{balescu}:
\begin{equation}
\mathrm{Im}R_x (\omega)=R_x''(\omega)=\frac{1}{\hbar}\tanh \left(\frac{\hbar \omega}{2k_B T}\right)C_x(\omega)\label{FDT}
\end{equation}
i.e.,
\begin{widetext}
\begin{equation}
C_x(\omega)=\frac{\hbar}{m}  \frac{\mathrm{Re}[K(\omega)]\left[\left(\left(\omega-\frac{\omega_0^2}{\omega}\right) + \mathrm{Im}[K(\omega)]\right)^{2}+\omega_{c}^{2}+\mathrm{Re}[K(\omega)]^{2} \right]\mathrm{coth}\left(\frac{\hbar\omega}{2k_{B}T} \right)}{\omega\left\lbrace  \left[\left(\left(\omega-\frac{\omega_0^2}{\omega}\right) + \mathrm{Im}[K(\omega)]\right)^{2}+\omega_{c}^{2}+\mathrm{Re}[K(\omega)]^{2}  \right]^{2}- 4\omega_c^{2}\left(\left(\omega-\frac{\omega_0^2}{\omega}\right) + \mathrm{Im}[K(\omega)]\right)^{2} \right\rbrace } \label{positioncorrelationomega}
\end{equation}
\end{widetext}

  The position autocorrelation function in the temporal domain, $C_x(t)$ can be obtained
via an inverse Fourier transform: 
\begin{equation}
    C_x(t)=\frac{1}{2 \pi }\int_{-\infty}^{\infty}C_x(\omega)e^{-i\omega t}d\omega
    \end{equation}
    Now, using Eq.(\ref{FDT}), one can get at finite temperatures,
    \begin{equation}
        C_x(t)=\frac{\hbar}{2\pi }\int_{-\infty}^{\infty} R_x''(\omega) coth\left(\frac{h\bar \omega}{2k_B T}\right)e^{-i \omega t}d\omega \label{positioncorrelationtime}
    \end{equation}
From the position autocorrelation function we obtain 
the position-velocity correlation function $C_{xv_x}(t)$ and 
the velocity autocorrelation function $C_{v_x}(t)$ by 
taking derivatives as follows:
\begin{equation}
C_{xv_x}(t)=\frac{d}{dt}C_x(t) \label{posvelcorr}
\end{equation}
and
\begin{equation}
 C_{v_x}(t)=-\frac{d^2}{dt^2}C_x(t)   \label{velautocorr}
\end{equation}
\section{A harmonically bound charged Brownian particle in a magnetic field}
\subsection{Ohmic Model}
\subsubsection{Long time tails in the position autocorrelation function}
Here we study and analyze the long time tails in the position autocorrelation function of a charged Brownian particle in a harmonic oscillator potential and a magnetic field and coupled to an Ohmic bath. \\
For an Ohmic model, the memory kernel in Eq.(\ref{qle}) is given by  \cite{Urbashione,surakaone}:
\begin{equation}
    K(\omega)=\gamma
\end{equation}
Using the above equation in Eq.(\ref{imresponsefunctionomega}) \cite{Urbashione,surakaone} we obtain the imaginary part ImR$_x(\omega)$ of the response function denoted by R$_x''(\omega)$: 
\begin{equation}
    R_x''(\omega)=\frac{\gamma}{m}\frac{\left(\left(\omega-\frac{\omega_0^2}{\omega}\right)^2+\omega_c^2+\gamma^2 \right)}{\omega\left[\left(\left(\omega-\frac{\omega_0^2}{\omega}\right)^2+\omega_c^2+\gamma^2 \right)^2-4 \omega_c^2\left(\omega-\frac{\omega_0^2}{\omega}\right)^2 \right]}\label{imrohmic}
\end{equation}
$R_x''(\omega)$ has poles at\\ $\omega=\frac{1}{2}(\pm \omega_c \pm i \gamma \pm \sqrt{4 \omega_0^2+\omega_c^2\pm 2i\omega_c \gamma-\gamma^2})$ and $ coth\left(\frac{\hbar \omega}{2k_B T}\right)$ has poles at $\omega=-in \pi  \Omega_{th}$ in the lower half plane, where $ \Omega_{th}=\frac{2  k_B T}{\hbar}$.

So from Eq.(\ref{positioncorrelationtime}), in the classical domain we find that $C_x(t)$ decays as e$^{-\gamma t}$. In contrast, 
in the quantum domain, i.e. for   
k$_BT<<\hbar\gamma,\omega_c$, the e$^{-\gamma t}$ fall off is much faster than the e$^{-n\pi  \Omega_{th} t}$ fall off\cite{graberttwo}. Hence the long time behaviour of $C_x(t)$ is determined by the poles of  $ coth\left(\frac{\hbar \omega}{2k_B T}\right)$. Accordingly, at large times, $C_x(t)$ is given by:
\begin{equation}
    C_x(t)=-i(\hbar \pi  \Omega_{th}/\pi )\sum_{n=1}^NR_x''(-in\pi  \Omega_{th})exp(-n \pi  \Omega_{th} t)
    \label{cothresidue}
\end{equation}
Here N=$\gamma/\pi  \Omega_{th}$ as the terms which decay faster than e$^{-\gamma t}$ are neglected.\\
Expanding $R_x''(-in\pi  \Omega_{th})$ in a Taylor series, we get,
\begin{equation}
\begin{split}
   &R_x''(-in\pi  \Omega_{th})=\frac{\gamma}{m}[\frac{(-in\pi  \Omega_{th})}{\omega_0^4}+ \\
   & \frac{(2\omega_0^2+3\omega_c^2-\gamma^2)(-in\pi  \Omega_{th})^3}{\omega_0^8}+ \\
  &  \frac{(3 \omega_0^4+5\omega_c^4-10\omega_c^2 \gamma^2+\gamma^4+4\omega_0^2(3\omega_c^2-\gamma^2))(-in\pi  \Omega_{th})^5}{\omega_0^{12}}+\\
   & O(-in\pi  \Omega_{th})^7 ] \label{chiinnu}
    \end{split}
\end{equation}
Therefore, from Eq.(\ref{cothresidue}) and Eq.(\ref{chiinnu}) we get, 
\begin{equation}
\begin{split}
  &  C_x(t)=-i\left(\frac{\hbar \pi  \Omega_{th}}{\pi } \right)\frac{\gamma}{m}\sum_{n=1}^N [\frac{(-in\pi  \Omega_{th})}{\omega_0^4}+ \\
& \frac{(2\omega_0^2+3\omega_c^2-\gamma^2)(-in\pi  \Omega_{th})^3}{\omega_0^8}+ \\
  &  \frac{(3 \omega_0^4+5\omega_c^4-10\omega_c^2 \gamma^2+\gamma^4+4\omega_0^2(3\omega_c^2-\gamma^2))(-in\pi  \Omega_{th})^5}{\omega_0^{12}}+ \\
  &O(-in\pi  \Omega_{th})^7 ] exp(-n \pi  \Omega_{th} t) \label{C_x(t)}
    \end{split}
\end{equation}
Thus, $C_x$(t) can be expressed as the sum of the contributions  from the different powers of (-in$\pi \Omega_{th}$):
\begin{equation}
    C_x(t)=C_{x}^{(1)}(t)+C_{x}^{(2)}(t)+C_{x}^{(3)}(t)+.....
\end{equation}
where
\begin{equation}
\begin{split}
    &C_{x}^{(1)}(t)=-i\left(\frac{\hbar \pi  \Omega_{th}}{\pi }\right)\frac{\gamma}{m}\frac{1}{\omega_0^4}\sum_{n=1}^N (-in\pi  \Omega_{th})\times\\& exp(-n \pi  \Omega_{th} t)   =\frac{\hbar \gamma \pi^2  \Omega_{th}^2}{\pi  m w_0^4} \frac{\partial}{\partial\zeta}\left[\frac{1}{exp(\zeta)-1}\right]\\ 
     \label{Cd1(t)}
    \end{split}
\end{equation}

\begin{equation}
\begin{split}
     &C_{x}^{(2)}(t)=-i\left(\frac{\hbar \pi  \Omega_{th}}{\pi }\right)\frac{\gamma}{m}\left(\frac{2\omega_0^2+3\omega_c^2-\gamma^2}{\omega_0^8}\right)\times \\
    & \sum_{n=1}^N (-in\pi  \Omega_{th})^3exp(-n \pi  \Omega_{th} t)\\
     =&-\frac{\hbar \pi  \Omega_{th}^4 \gamma}{\pi  m}\left(\frac{2\omega_0^4+3\omega_c^2-\gamma^2}{\omega_0^8}\right) \frac{\partial^3}{\partial \zeta^3}\left[\frac{1}{exp(\zeta)-1}\right] \label{Cd2(t)}
     \end{split}
\end{equation}
\begin{equation}
\begin{split}
   &  C_{x}^{(3)}(t)=-i\left(\frac{\hbar \pi  \Omega_{th}}{\pi }\right)\frac{\gamma}{m} \\ &\left(\frac{3\omega_0^4+5\omega_c^4-10\omega_c^2\gamma^2+\gamma^4+4\omega_0^2(3\omega_c^2-\gamma^2)}{\omega_0^{12}}\right) \\
    & \sum_{n=1}^N (-in\pi  \Omega_{th})^5exp(-n \pi  \Omega_{th} t)= \\ 
    & \frac{\hbar \pi^6  \Omega_{th}^6 \gamma}{\pi  m}\left(\frac{3\omega_0^4+5\omega_c^4-10\omega_c^2\gamma^2+\gamma^4+4\omega_0^2(3\omega_c^2-\gamma^2)}{\omega_0^{12}}\right)\\
   & \frac{\partial^5}{\partial \zeta^5}\left[\frac{1}{exp(\zeta)-1}\right] \label{Cd3(t)}
 \end{split}    
\end{equation}
$\zeta$ being equal to $\pi  \Omega_{th} t$.

\subsubsection{Long time tails in the Position-Velocity Correlation Function}
The position-velocity correlation function can be determined from the position autocorrelation function using Eq.(\ref{posvelcorr}):
\begin{equation}
    C_{xv_x}(t)=\frac{d}{dt}C_x(t)
=C_{xv_x}^{(1)}(t)+C_{xv_x}^{(2)}(t)+C_{xv_x}^{(3)}(t)+....
\end{equation}
where
\begin{equation}
C_{xv_x}^{(1)}(t)=\frac{d}{dt}C_{x}^{(1)}(t)
\end{equation}
\begin{equation}
C_{xv_x}^{(2)}(t)=\frac{d}{dt}C_{x}^{(2)}(t) 
\end{equation}

\begin{equation}
C_{xv_x}^{(3)}(t)=\frac{d}{dt}C_{x}^{(3)}(t)
\end{equation}

Therefore, using eqs.(\ref{Cd1(t)}-\ref{Cd3(t)}), we get,
\begin{equation}
\begin{split}
    C_{xv_x}^{(1)}(t)=
   & \frac{\hbar \gamma \pi^3  \Omega_{th}^3}{\pi  m \omega_0^4}\frac{\partial^2}{\partial \zeta^2}\left[\frac{1}{exp(\zeta)-1}\right]\\
    \end{split}
\end{equation}

\begin{equation}
\begin{split}
    C_{xv_x}^{(2)}(t)=
   & -\frac{\hbar\pi^5  \Omega_{th}^5 \gamma}{\pi  m}\left(\frac{2\omega_0^2+3\omega_c^2-\gamma^2}{\omega_0^8}\right) \times\\
   &\frac{\partial^4}{\partial \zeta^4}\left[\frac{1}{exp(\zeta)-1}\right]\\
   \end{split}
\end{equation}
\begin{equation}
\begin{split}
  C_{xv_x}^{(3)}(t) =&\frac{\hbar \pi^7  \Omega_{th}^7 \gamma}{\pi  m}\times\\
  &\left(\frac{3\omega_0^4+5\omega_c^4-10\omega_c^2\gamma^2+\gamma^4+4\omega_0^2(3\omega_c^2-\gamma^2)}{\omega^{12}}\right)\\
 &\times \frac{\partial^6}{\partial \zeta^6}\left[\frac{1}{exp(\zeta)-1}\right]
  \end{split}
\end{equation}

\subsubsection{Long time tails in the Velocity Autocorrelation Function}
The velocity autocorrelation function is determined from the position autocorrelation using Eq.(\ref{velautocorr})
\begin{equation}
C_{v_x}(t)=-\frac{d^2}{dt^2}C_x(t)=C_{v_x}^{(1)}(t)+C_{v_x}^{(2)}(t)+C_{v_x}^{(3)}(t)+...
\end{equation}
where,
\begin{equation}
C_{v_x}^{(1)}(t)=-\frac{d^2}{dt^2}C_{x}^{(1)}(t)
\end{equation}
\begin{equation}
C_{v_x}^{(2)}(t)=-\frac{d^2}{dt^2}C_{x}^{(2)}(t)
\end{equation}
\begin{equation}
C_{v_x}^{(3)}(t)=-\frac{d^2}{dt^2}C_{x}^{(3)}(t)
\end{equation}

As in the case of the position velocity correlation, in this case too, using, Eqs.(\ref{Cd1(t)}-\ref{Cd3(t)}), we get,

\begin{equation}
\begin{split}
    C_{v_x}^{(1)}(t)&=-\frac{\hbar \pi^4  \Omega_{th}^4 \gamma}{\pi  m \omega_0^4}\frac{\partial^3}{\partial \zeta^3}\left[\frac{1}{exp(\zeta)-1}\right]
    \end{split}
\end{equation}
\begin{equation}
\begin{split}
    C_{v_x}^{(2)}(t)=\frac{\hbar\pi^6 \Omega_{th}^6 \gamma }{\pi  m} \left(\frac{2\omega_0^2+3\omega_c^2-\gamma^2}{\omega_0^8}\right)\frac{\partial^5}{\partial \zeta^5}\left[\frac{1}{exp(\zeta)-1} \right]
   \end{split}
    \end{equation}
\begin{equation}
\begin{split}
  C_{v_x}^{(3)}(t)
 =&-\frac{\hbar \pi^8  \Omega_{th}^8 \gamma}{\pi  m} \times\\ 
 &\left(\frac{3\omega_0^4+5\omega_c^4-10\omega_c^2\gamma^2+\gamma^4+4\omega_0^2(3\omega_c^2-\gamma^2)}{\omega^{12}}\right)\\
 &\times \frac{\partial^7}{\partial \zeta^7}\left[\frac{1}{exp(\zeta)-1}\right]
  \end{split}
\end{equation}
\subsection{Drude Model}
\subsubsection{Long time tails in the position autocorrelation Function}
For the Drude Model, the real and imaginary parts
\footnote{$K'$ and $K''$ stands for the $Re$ part and the $Im$ part of $K$ respectively, consistent with the notation $R_{x}''$ used for $Im(R_{x})$.} of the memory Kernel are respectively given by \cite{surakaone}:
\begin{align}
    K'(\omega)=\frac{\gamma}{1+\omega^2 \tau^2} \label{drudememkernelreal}\\
    K''(\omega)=\frac{\omega \gamma \tau}{1+\omega^2 \tau^2}
         \label{drudememkernelim}
\end{align}

\begin{widetext}
Using the above equation in Eq.(\ref{imresponsefunctionomega}) \cite{surakaone} we obtain :

\begin{equation}
    R_x''(\omega)=\frac{\gamma \left(\omega_c^2+\frac{\gamma^2}{(1+\omega^2 \tau^2)^2}+\left((\omega-\frac{\omega_0^2}{\omega})+\frac{\omega \gamma \tau}{(1+\omega^2 \tau^2)}\right)^2\right)}{m\omega(1+\omega^2 \tau^2)\left(-4 \omega_c^2 \left((\omega-\frac{\omega_0^2}{\omega})+\frac{\omega \gamma \tau}{1+\omega^2 \tau^2} \right)^2 +\left(\omega_c^2+\frac{\gamma^2}{(1+\omega^2 \tau^2)^2}+\left((\omega-\frac{\omega_0^2}{\omega})+\frac{\omega \gamma \tau}{1+\omega^2 \tau^2} \right)^2 \right)^2 \right)}
\end{equation} \\
\end{widetext}
$R_x''(\omega)$ has poles at $\frac{1}{2}\left(-\omega_c-\frac{i}{\tau}+\frac{\sqrt{-1-2i\omega_c\tau-4\gamma\tau+\omega_c^2 \tau^2}}{\tau} \right)$, $\frac{1}{2}\left(\omega_c+\frac{i}{\tau}+\frac{\sqrt{-1-2i\omega_c\tau-4\gamma\tau+\omega_c^2 \tau^2}}{\tau} \right)$, $\frac{1}{2}\left(\omega_c-\frac{i}{\tau}-\frac{\sqrt{-1+2i\omega_c\tau-4\gamma\tau+\omega_c^2 \tau^2}}{\tau} \right)$, $\frac{1}{2}\left(-\omega_c+\frac{i}{\tau}-\frac{\sqrt{-1+2i\omega_c\tau-4\gamma\tau+\omega_c^2 \tau^2}}{\tau} \right)$ in the lower half plane. The imaginary parts of the poles that contribute to the decay of the correlation functions are $\big(\pm \frac{1}{2 \tau}-\\
\frac{(4 \omega_c^2 \tau^2+(-1-4\gamma \tau+\omega_c^2 \tau^2)^2)^{1/4}sin[\frac{1}{2}Arg[-1\pm2i\omega_c \tau-4 \gamma \tau+\omega_c^2 \tau^2]]}{2\tau}\big)$. As in the Ohmic case, in the quantum domain, i.e. for   
k$_BT<<\hbar \gamma,\omega_c$,$\frac{1}{\tau}$, the   e$^{(-1/2\tau) t}$ fall off is more rapid compared to the  $e^{-n\pi  \Omega_{th} t}$ fall off and the long time behaviour of $C_x(t)$ is determined by the poles of  $ coth\left(\frac{\hbar \omega}{2k_B T}\right)$. Accordingly, at large times, $C_x(t)$ is given by: 
\begin{equation}
    C_x(t)=-i(\hbar \pi  \Omega_{th}/\pi )\sum_{n=1}^NR_x''(-in\pi  \Omega_{th})exp(-n \pi  \Omega_{th} t)
    \label{cothresiduedrude}
\end{equation}
\hspace*{1cm}Here N=$\gamma/\pi  \Omega_{th}$ as the terms which decay faster than e$^{-\gamma t}$ are neglected. \\ 
Expanding $R_x''(-in\pi  \Omega_{th})$ in a Taylor series, we get,
\begin{equation}
\begin{split}
 R_x''(-in\pi  \Omega_{th})&=\frac{\gamma}{m}\Big[A_1(-in\pi  \Omega_{th})+A_2(-in\pi  \Omega_{th})^3+\\
 &A_3(-in\pi  \Omega_{th})^5 \Big] \label{chiinnudrude}
 \end{split}
    \end{equation}
where,
\begin{equation}
    A_1=1/\omega_0^4
\end{equation}
\begin{equation}
    A_2=(2\omega_0^2+3\omega_c^2-\gamma^2+2\omega_0^2\gamma \tau+\omega_0^4\tau^2)/\omega_0^8
\end{equation}

\begin{equation}
\begin{split}
   &A_3= (3\omega_0^4+12\omega_0^2\omega_c^2+5\omega_c^4-4\omega_0^2\gamma^2-10\omega_c^2\gamma^2+\gamma^4+\\
  & 6\omega_0^4\gamma \tau+12\omega_0^2\omega_c^2\gamma \tau-4\omega_0^2\gamma^3\tau-2\omega_0^6\tau^2-3\omega_0^4\omega_c^2\tau^2+ \\& 6\omega_0^4\gamma^2\tau^2 
   -4\omega_0^6\gamma\tau^3+\omega_0^8\tau^4)/\omega_0^{12}
   \end{split}
\end{equation}
Therefore, from Eq.(\ref{cothresiduedrude}) and Eq.(\ref{chiinnudrude}) we get, 
\begin{equation}
\begin{split}
    &C_x(t)=-i\left(\frac{\hbar \pi  \Omega_{th}}{\pi }\right)\frac{\gamma}{m}\sum_{n=1}^N\Big[A_1(-in\pi  \Omega_{th})^{-1}+\\
    &A_2(-in\pi  \Omega_{th})+A_3(-in\pi  \Omega_{th})^3\Big]exp(-n \pi  \Omega_{th} t) \label{C_x(t)drude}
    \end{split}
\end{equation}

Thus, 
\begin{equation}
    C_x(t)=C_{x}^{(1)}(t)+C_{x}^{(2)}(t)+C_{x}^{(3)}(t)+.....
\end{equation}
where
\begin{equation}
\begin{split}
   & C_{x}^{(1)}(t)=-i\left(\frac{\hbar \pi  \Omega_{th}}{\pi }\right)\frac{\gamma}{m}A_1\times \\
   & \sum_{n=1}^N(-in\pi  \Omega_{th})exp(-n \pi  \Omega_{th} t)=\\
   &\frac{\hbar \gamma \pi^2  \Omega_{th}^2}{\pi  m}A_1 \frac{\partial}{\partial \zeta}\left[\frac{1}{exp(\zeta)-1}\right] \label{J1(t)}
    \end{split}
\end{equation}

\begin{equation}
\begin{split}
     &C_{x}^{(2)}(t)=-i\left(\frac{\hbar \pi  \Omega_{th}}{\pi }\right)\frac{\gamma}{m}A_2 \times \\
    & \sum_{n=1}^N (-in\pi  \Omega_{th})^3exp(-n \pi  \Omega_{th} t)=\\&-\frac{\hbar \pi^4  \Omega_{th}^4 \gamma}{\pi  m}A_2\frac{\partial^3}{\partial \zeta^3}\left[\frac{1}{exp(\zeta)-1}\right] \label{J2(t)}
     \end{split}
\end{equation}

\begin{equation}
\begin{split}
    & C_{x}^{(3)}(t)=-i\left(\frac{\hbar \pi  \Omega_{th}}{\pi }\right)\frac{\gamma}{m}A_3\times\\
    &\sum_{n=1}^N (-in\pi  \Omega_{th})^5exp(-n \pi  \Omega_{th} t)= \\
     &\frac{\hbar \pi^6  \Omega_{th}^6 \gamma}{\pi  m}A_3\frac{\partial^5}{\partial \zeta^5}\left[\frac{1}{exp(\zeta)-1}\right] \label{J3(t)}
 \end{split}    
\end{equation}
$\zeta$ being equal to $\pi  \Omega_{th} t$.
\subsubsection{Long time tails in the Position-Velocity Correlation Function}
The position-velocity correlation function can be determined from the position autocorrelation function using Eq.(\ref{posvelcorr}):
\begin{align}
    C_{xv_x}(t)=\frac{d}{dt}C_x(t)
=C_{xv_x}^{(1)}(t)+C_{xv_x}^{(2)}(t)+C_{xv_x}^{(3)}(t)+...\end{align}
where
\begin{align}
C_{xv_x}^{(1)}(t)=\frac{d}{dt}C_{x}^{(1)}(t)\\
C_{xv_x}^{(2)}(t)=\frac{d}{dt}C_{x}^{(2)}(t) \\
C_{xv_x}^{(3)}(t)=\frac{d}{dt}C_{x}^{(3)}(t)
\end{align}


Therefore, using Eqs.(\ref{J1(t)}-\ref{J3(t)}), we get,
\begin{equation}
\begin{split}
    C_{xv_x}^{(1)}(t)=\frac{\hbar \gamma \pi^3  \Omega_{th}^3}{\pi  m }A_1\frac{\partial^2}{\partial \zeta^2}\left[\frac{1}{exp(\zeta)-1}\right]
    \end{split}
\end{equation}

\begin{equation}
\begin{split}
    C_{xv_x}^{(2)}(t)=-\frac{\hbar\pi^4  \Omega_{th}^4 \gamma}{\pi  m}A_2\frac{\partial^4}{\partial \zeta^4}\left[\frac{1}{exp(\zeta)-1}\right]
    \end{split}
\end{equation}
\begin{equation}
\begin{split}
  C_{xv_x}^{(3)}(t)=\frac{\hbar \pi^7  \Omega_{th}^7 \gamma}{\pi  m}A_3\frac{\partial^6}{\partial \zeta^6}\left[\frac{1}{exp(\zeta)-1}\right]
  \end{split}
\end{equation}


\subsubsection{Long time tails in the Velocity Autocorrelation Function}
The velocity autocorrelation function is determined from the position autocorrelation using Eq.(\ref{velautocorr})
\begin{equation}
C_{v_x}(t)=-\frac{d^2}{dt^2}C_x(t)=C_{v_x}^{(1)}(t)+C_{v_x}^{(2)}(t)+C_{v_x}^{(3)}(t)+...
\end{equation}
where,
\begin{align}
C_{v_x}^{(1)}(t)=-\frac{d^2}{dt^2}C_{x}^{(1)}(t)\\
C_{v_x}^{(2)}(t)=-\frac{d^2}{dt^2}C_{x}^{(2)}(t) \\
C_{v_x}^{(3)}(t)=-\frac{d^2}{dt^2}C_{x}^{(3)}(t)
\end{align}

As in the Ohmic case, using Eqs.(\ref{J1(t)}-\ref{J3(t)}), we get,

\begin{equation}
\begin{split}
    C_{v_x}^{(1)}(t)=-\frac{\hbar \pi^4  \Omega_{th}^4 \gamma}{\pi  m}A_1\frac{\partial^3}{\partial \zeta^3}\left[\frac{1}{exp(\zeta)-1}\right]
    \end{split}
\end{equation}

\begin{equation}
\begin{split}
    C_{v_x}^{(2)}(t)=\frac{\hbar\pi^6  \Omega_{th}^6 \gamma }{\pi  m}A_2\frac{\partial^5}{\partial \zeta^5}\left[\frac{1}{exp(\zeta)-1}\right]
    \end{split}
\end{equation}

\begin{equation}
\begin{split}
  C_{v_x}^{(3)}(t)=-\frac{\hbar \pi^8  \Omega_{th}^8 \gamma}{\pi  m}A_3\frac{\partial^7}{\partial \zeta^7}\left[\frac{1}{exp(\zeta)-1}\right]
  \end{split}
\end{equation}

Here, we have analysed the following three cases: \\ 
(i) Overdamped case ($\omega_c <<\gamma$);\\ 
(ii) Critically damped case ($\omega_c \sim \gamma $);\\
(iii) Underdamped case ($\omega_c >> \gamma$)\\ 
 for the position autocorrelation, the position-velocity correlation and the velocity autocorrelation and the plots  for various oscillatory and damping rates are presented in Sec IV.
The analysis of the correlation functions show that at zero temperature, the long time decay of the correlation functions
exhibit power law trends as shown in Table I and II.  In contrast, at finite temperatures, the correlation functions decay exponentially. For instance, for the position correlation function one gets the following long time exponentially decaying behaviour:
 \begin{align}
 &C_x^{(1)}(t) \approx \frac{\hbar\gamma\pi^2\Omega_{th}^2}{\pi m \omega_0^4}exp(-\zeta)\\
 &C_x^{(2)}(t)\approx -\frac{\hbar \pi \Omega_{th}^4 \gamma}{\pi m} \left(\frac{2\omega_0^4+3\omega_c^2-\gamma^2}{\omega_0^8} \right) exp(-\zeta)
 \end{align}\\
 We end this sub-section with the two tables, where we display all the power law behaviours for the correlation functions for the Ohmic (Table I) and the Drude (Table II) bath models.
 
 \begin{widetext}
\begin{center}
\begin{table}[H]
\caption{ Table for power law tails for different correlation functions for the Ohmic bath}
\begin{tabular}{ |p{2cm}|p{3.8cm}|p{4.2cm}|p{3.7cm}| p{3.7cm}| }

 \hline
 \multicolumn{5}{|c|}{Power law tails for the Ohmic bath} \\
 \hline
\centering{Parameters} &  \centering{$\omega_c =\omega_0 =0$}  & \centering{$\omega_c\neq 0,\omega_0 =0$} &  \centering{$\omega_c=0, \omega_0 \neq 0$} & \centering{$\omega_c\neq0,\omega_0 \neq 0$}\cr
\hline
 \centering{$C_x(t)$}  & \centering{$-\frac{\hbar}{\pi  m \gamma}\log\left(\pi  \Omega_{th} t \right) 
 +\frac{\hbar}{\pi m \gamma^3}t^{-2}$} & \centering{$-\frac{\hbar \gamma }{\pi  m}\alpha_1(\omega_c,\gamma) \log\left(\pi  \Omega_{th} t \right)- \frac{\hbar \gamma}{\pi m}\alpha_{2}(\omega_c,\gamma,0)t^{-2}$} & \centering{$-\frac{\hbar \gamma}{\pi m }A_1 (\omega_0) t^{-2}+\frac{6\hbar \gamma}{\pi m} A_{2}(\omega_0,0,\gamma,0) t^{-4}$}  &\centering{$-\frac{\hbar \gamma}{\pi m }A_1 (\omega_0) t^{-2}+\frac{6\hbar \gamma}{\pi m} A_{2}(\omega_0,\omega_c,\gamma,0)t^{-4}$}  \cr
 \hline
 \centering{$C_{xv_x}(t)$} & \centering{$-\frac{\hbar}{\pi m \gamma}t^{-1}-\frac{2\hbar}{\pi m \gamma^3}t^{-3} $}  & \centering{- $\frac{\hbar \gamma}{\pi m } \alpha_1(\omega_c,\gamma) t^{-1}+\frac{2\hbar \gamma}{\pi m} \alpha_{2}(\omega_c,\gamma,0) t^{-3}$} & \centering{$\frac{2 \hbar \gamma}{\pi m }A_1 (\omega_0)t^{-3}-\frac{24 \hbar \gamma}{\pi m}A_{2}(\omega_0,0,\gamma,0)t^{-5}$} & \centering{$\frac{2 \hbar \gamma}{\pi m }A_1 (\omega_0) t^{-3}-\frac{24 \hbar \gamma}{\pi m} A_{2}(\omega_0,\omega_c,\gamma,0)t^{-5}$} \cr
 \hline
  \centering{$C_{v_x}(t)$} &\centering{$-\frac{\hbar}{\pi m \gamma}t^{-2}-\frac{6\hbar}{\pi m \gamma^3}t^{-4} $}  &  \centering{$-\frac{\hbar \gamma}{\pi m }\alpha_1(\omega_c,\gamma) t^{-2}+\frac{6 \hbar \gamma}{\pi m}\alpha_{2}(\omega_c,\gamma,0) t^{-4} $} & \centering{$\frac{6 \hbar \gamma}{\pi m }A_1(\omega_0)t^{-4}-\frac{120 \hbar \gamma}{\pi m}A_{2}(\omega_0,0,\gamma,0) t^{-6}$} &\centering{$\frac{6 \hbar \gamma}{\pi m }A_1(\omega_0)t^{-4}-\frac{120 \hbar \gamma}{\pi m} A_{2}(\omega_0,\omega_c,\gamma,0)t^{-6}$}\cr
  \hline
\end{tabular}

\label{table1}
\end{table}

\begin{table}[H]
\caption{ Table for power law tails for different correlation functions for the Drude bath}
\begin{tabular}{ |p{2cm}|p{3.8cm}|p{4.2cm}|p{3.7cm}| p{3.7cm}| }

 \hline
 \multicolumn{5}{|c|}{Power law tails for the Drude bath} \\
 \hline
\centering{Parameters} &  \centering{$\omega_c =\omega_0 =0$}  & \centering{$\omega_c\neq 0,\omega_0 =0$} &  \centering{$\omega_c=0, \omega_0 \neq 0$} & \centering{$\omega_c\neq0,\omega_0 \neq 0$}\cr
\hline
 \centering{$C_x(t)$}  & \centering{$-\frac{\hbar}{\pi  m \gamma}\log\left(\pi  \Omega_{th} t \right) 
 +\frac{\hbar}{\pi m \gamma^3}t^{-2}$} & \centering{$-\frac{\hbar \gamma }{\pi  m}\alpha_1(\omega_c,\gamma) \log\left(\pi  \Omega_{th} t \right)- \frac{\hbar \gamma}{\pi m}\alpha_{2}(\omega_c,\gamma,\tau)t^{-2}$} & \centering{$-\frac{\hbar \gamma}{\pi m }A_1(\omega_0)t^{-2}+\frac{6\hbar \gamma}{\pi m} A_{2}(\omega_0,0,\gamma,\tau) t^{-4}$}  &\centering{$-\frac{\hbar \gamma}{\pi m }A_1(\omega_0)t^{-2}+\frac{6\hbar \gamma}{\pi m} A_{2}(\omega_0,\omega_c,\gamma,\tau)t^{-4}$} \cr
 \hline
 \centering{$C_{xv_x}(t)$} & \centering{$-\frac{\hbar}{\pi m \gamma}t^{-1}-\frac{2\hbar}{\pi m \gamma^3}t^{-3} $}  & \centering{- $\frac{\hbar \gamma}{\pi m } \alpha_1(\omega_c,\gamma) t^{-1}+\frac{2\hbar \gamma}{\pi m} \alpha_{2}(\omega_c,\gamma,\tau) t^{-3}$} & \centering{$\frac{2 \hbar \gamma}{\pi m }A_1(\omega_0)t^{-3}-\frac{24 \hbar \gamma}{\pi m}A_{2}(\omega_0,0,\gamma,\tau)t^{-5}$} & \centering{$\frac{2 \hbar \gamma}{\pi m }A_1(\omega_0)t^{-3}-\frac{24 \hbar \gamma}{\pi m} A_{2}(\omega_0,\omega_c,\gamma,\tau)t^{-5}$} \cr
 \hline
  \centering{$C_{v_x}(t)$} &\centering{$-\frac{\hbar}{\pi m \gamma}t^{-2}-\frac{6\hbar}{\pi m \gamma^3}t^{-4} $}  &  \centering{$-\frac{\hbar \gamma}{\pi m }\alpha_1(\omega_c,\gamma) t^{-2}+\frac{6 \hbar \gamma}{\pi m}\alpha_{2}(\omega_c,\gamma,\tau) t^{-4} $} & \centering{$\frac{6 \hbar \gamma}{\pi m }A_1(\omega_0)t^{-4}-\frac{120 \hbar \gamma}{\pi m}A_{2}(\omega_0,0,\gamma,\tau) t^{-6}$} &\centering{$\frac{6 \hbar \gamma}{\pi m }A_1(\omega_0)t^{-4}-\frac{120 \hbar \gamma}{\pi m} A_{2}(\omega_0,\omega_c,\gamma,\tau)t^{-6}$}\cr
  \hline
  
 \end{tabular}\\ \\

\label{table1}
\caption*{\textbf{Note:} In the above tables, 
   $\alpha_1(\omega_c,\gamma)=\frac{1}{\omega_c^2+\gamma^2}$, $\alpha_2(\omega_c,\gamma,\tau)=\frac{3\omega_c^2-\gamma^2+6\omega_c^2\gamma\tau+2\gamma^3\tau+\omega_c^4\tau^2-3\omega_c^2\gamma^2\tau^2}{(\omega_c^2+\gamma^2)^3}$ and the expressions for $A_1$, $A_2$ are mentioned in the text in Sec. $III$.}
\end{table}
\end{center}
\vspace*{-1cm}

\begin{figure}[H]
\centering
\includegraphics[width=1.02\textwidth]{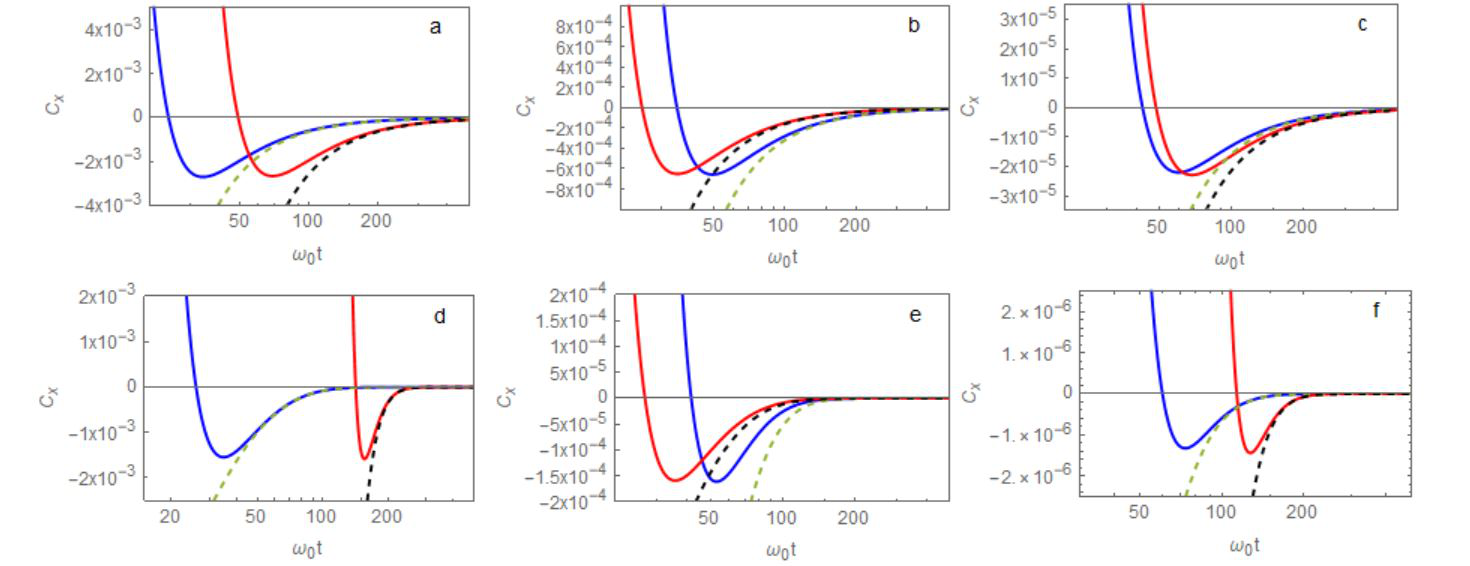} 
\caption{The position autocorrelation function ($C_x$) vs. $\omega_0 t$ in log-linear scale (i) at a very low temperature ($\pi \Omega_{th}=10^{-4}$) : (a) Over-damped ($\omega_0$=2, $\omega_c$=10, $\gamma$=20, $\pi \Omega_{th}=10^{-4}$); (b)  Critically-damped ($\omega_0$=2, $\omega_c$=10, $\gamma$=10, $\pi \Omega_{th}=10^{-4}$);  (c) Under-damped ($\omega_0$=2, $\omega_c$=10, $\gamma$=0.5, $\pi \Omega_{th}=10^{-4}$) and (ii) at a finite temperature ($\pi \Omega_{th}$=0.05) :  (d) Over-damped ($\omega_0$=2, $\omega_c$=10, $\gamma$=20, $\pi \Omega_{th}$=0.05); (e) Critically-damped ($\omega_0$=2, $\omega_c$=10, $\gamma$=10, $\pi \Omega_{th}$=0.05); (f) Under-damped ($\omega_0$=2, $\omega_c$=10, $\gamma$=0.5, $\pi \Omega_{th}$=0.05)  [The blue solid curve denotes the Ohmic and the red solid curve denotes the Drude case ($\tau=12$). The green and black dashed curves for  figs.(a-c) represent the power law decays for the Ohmic and Drude case respectively, as given in the tables. The dashed curves in the finite temperature plots (figs.(d-f)) represent the exponential decays for the corresponding Ohmic and Drude cases]}\label{fig1}
\end{figure}

\begin{figure}[H]
\centering
\includegraphics[width=1\textwidth]{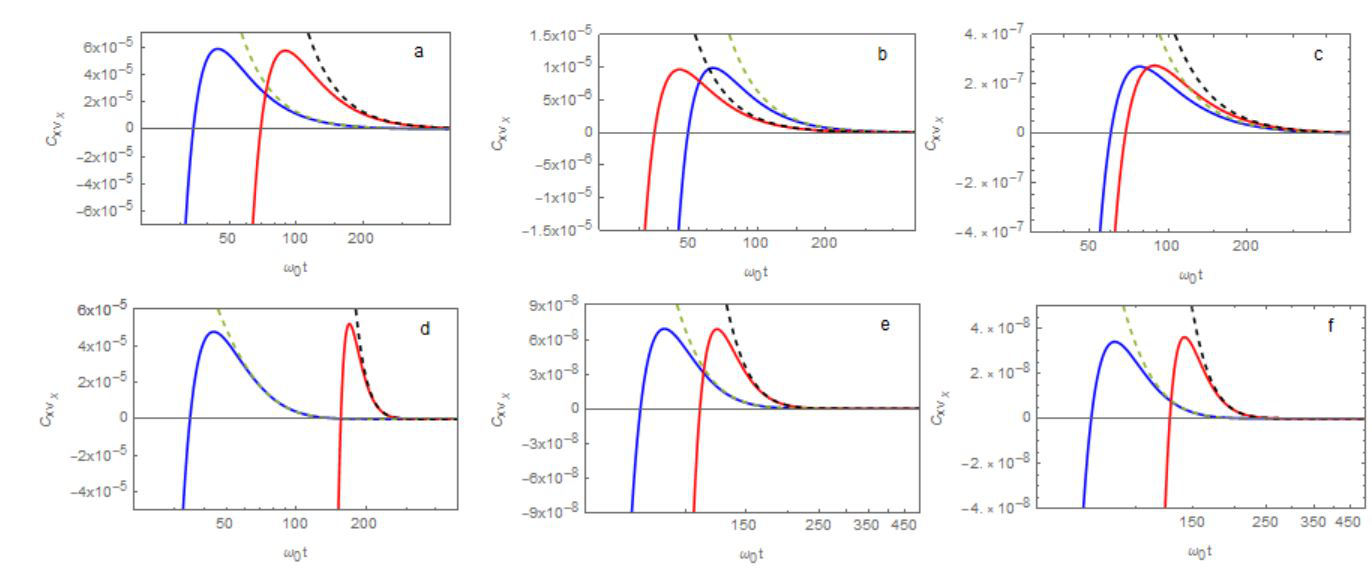} 

\caption{The position-velocity correlation function ($C_{xv_x}$) vs. $\omega_0 t$ in log-linear scale (i) at a very low temperature ($\pi \Omega_{th}=10^{-4}$) : (a) Over-damped ($\omega_0$=2, $\omega_c$=10, $\gamma$=20, $\pi \Omega_{th}=10^{-4}$); (b)  Critically-damped ($\omega_0$=2, $\omega_c$=10, $\gamma$=10, $\pi \Omega_{th}=10^{-4}$);  (c) Under-damped ($\omega_0$=2, $\omega_c$=10, $\gamma$=0.5, $\pi \Omega_{th}=10^{-4}$) and (ii) at a finite temperature ($\pi \Omega_{th}$=0.05) :  (d) Over-damped ($\omega_0$=2, $\omega_c$=10, $\gamma$=20, $\pi \Omega_{th}$=0.05); (e) Critically-damped ($\omega_0$=2, $\omega_c$=10, $\gamma$=10, $\pi \Omega_{th}$=0.05); (f) Under-damped ($\omega_0$=2, $\omega_c$=10, $\gamma$=0.5, $\pi \Omega_{th}$=0.05)  [The blue solid curve denotes the Ohmic and the red solid curve denotes the Drude case ($\tau=12$). The green and black dashed curves for  figs.(a-c) represent the power law decays for the Ohmic and Drude case respectively, as given in the tables. The dashed curves in the finite temperature plots (figs.(d-f)) represent the exponential decays for the corresponding Ohmic and Drude cases]}\label{fig2}
\end{figure}

\begin{figure}[H]
\centering
\includegraphics[width=1.02\textwidth]{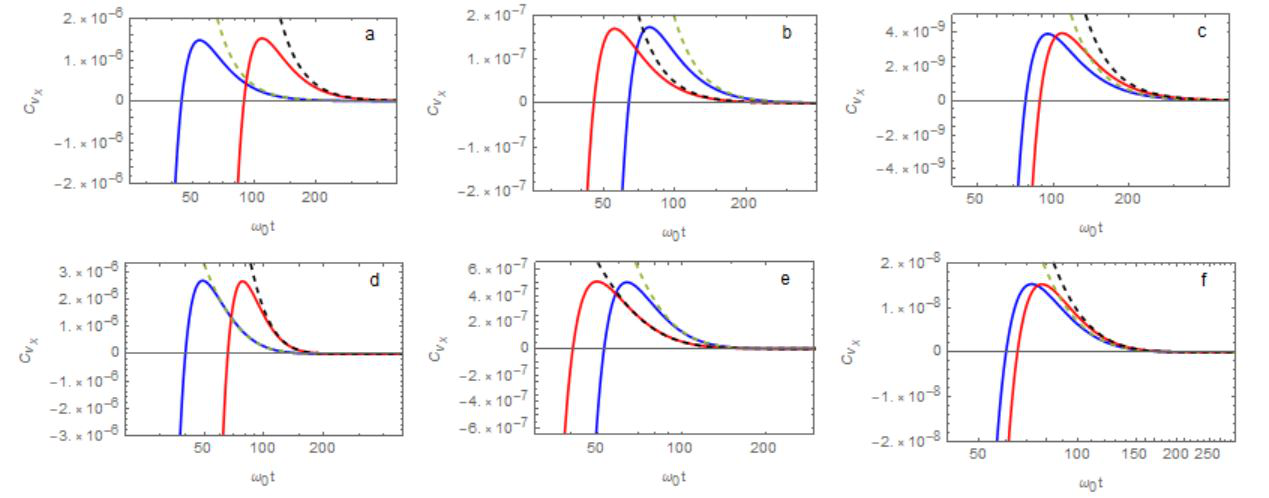} 

\caption{The velocity autocorrelation function ($C_{v_x}$) vs. $\omega_0 t$ in log-linear scale (i) at a very low temperature ($\pi \Omega_{th}=10^{-4}$) : (a) Over-damped ($\omega_0$=2, $\omega_c$=10, $\gamma$=20, $\pi \Omega_{th}=10^{-4}$); (b)  Critically-damped ($\omega_0$=2, $\omega_c$=10, $\gamma$=10, $\pi \Omega_{th}=10^{-4}$);  (c) Under-damped ($\omega_0$=2, $\omega_c$=10, $\gamma$=0.5, $\pi \Omega_{th}=10^{-4}$) and (ii) at a finite temperature ($\pi \Omega_{th}$=0.05) :  (d) Over-damped ($\omega_0$=2, $\omega_c$=10, $\gamma$=20, $\pi \Omega_{th}$=0.05); (e) Critically-damped ($\omega_0$=2, $\omega_c$=10, $\gamma$=10, $\pi \Omega_{th}$=0.05); (f) Under-damped ($\omega_0$=2, $\omega_c$=10, $\gamma$=0.5, $\pi \Omega_{th}$=0.05)  [The blue solid curve denotes the Ohmic and the red solid curve denotes the Drude case ($\tau=12$). The green and black dashed curves for  figs.(a-c) represent the power law decays for the Ohmic and Drude case respectively, as given in the tables. The dashed curves in the finite temperature plots (figs.(d-f)) represent the exponential decays for the corresponding Ohmic and Drude cases]}\label{fig3}
\end{figure}

\end{widetext}

\section{Results and Discussion}

Here, we have plotted the position autocorrelation, position-velocity correlation and velocity autocorrelation functions for the Ohmic and Drude models, exhibiting a power law behaviour at very low temperatures (near zero) and exponential behaviour at finite temperatures. The plots  are done for various damping rates representing the over-damped, critically damped and under-damped regimes respectively.\\
\hspace*{0.2cm} Figs.(1-3) show that, at finite temperatures, the position autocorrelation function, position-velocity correlation and velocity autocorrelation functions ($C_x$, $C_{xv_x}$ and $C_{v_x}$) decay exponentially, however, when the temperature approaches zero, the exponential decays go over to power law decays with various power law exponents for the different correlation functions as shown in Table-I and Table-II. We notice that a harmonic confinement leads to 
a faster decay of the long time tails. The presence of the cyclotron frequency leads to a modification of the coefficient
of the power law tails. \\
\hspace*{0.2cm}The plots show, as expected, that the curves fall off faster (and eventually settle down to the long time power law or exponential  trend) in the overdamped case compared to the underdamped case, with the critically damped case exhibiting an intermediate behaviour between these two extreme cases.
Furthermore, one can notice from Table-II that the first terms of the correlation functions are independent of the 
Drude time $\tau$ .
\hspace*{0.2cm} Therefore, in order to glean out the dependence of the 
long time tails on the Drude time $\tau$ in the case of the Drude bath, we have retained the first two terms both in the finite temperature and zero temperature analyses and plots. While the exponents of the power law tails are the same for the Ohmic and 
Drude baths, the coefficients of the power law tails are distinct in the two cases. Thus, the Drude time does play a role in controlling the coefficients of the power law tails and thus affects the overall trends of decay of the correlation functions. \\
\hspace*{0.2cm}The results for the long time behaviours of the  position correlation, position-velocity correlation and velocity autocorrelation function discussed in this paper 
are entirely new in the context of the quantum Brownian motion of a charged particle in the presence of a harmonic oscillator potential and a magnetic field. 
\section{Conclusion}

In this paper we have studied in detail the long time tail behaviour of the position autocorrelation function, the position-velocity correlation function and the velocity autocorrelation function of a charged particle in a magnetic field in a viscous medium.\\
\hspace*{0.2cm}As in the case of a neutral particle \cite{graberttwo} we see a transition 
from an exponentially damped trend to a power law behaviour
as one goes from a finite temperature to zero temperature. 
We get a variety of power law trends in the overdamped, 
underdamped and critically damped regimes which explore
the relative strengths of the cyclotron frequency, the trapping frequency and the viscous damping. We display our results in tabular form and also plot them to visually study the trends. Our results converge to the previously 
researched neutral particle behaviour\cite{graberttwo} for the long time tails in the appropriate limits and go far beyond that in analysing in detail the long time tail trends for the position autocorrelation, the position-velocity correlation and the velocity autocorrelation function for a charged particle in a magnetic field in a 
viscous medium. 
 Also, earlier researchers who had studied the  
long time tails of a neutral particle\cite{graberttwo} had confined their analysis to 
the context of an Ohmic bath. We do our analysis both for the Ohmic and the Drude baths and notice how the amplitude of the 
long time tail gets modified by the presence of the Drude 
time scale $\tau$. Thus, our long time tail analysis for the 
Drude model goes beyond existing literature both for a neutral particle and a charged particle in a viscous medium
in the presence of a harmonic potential. We notice that 
the presence of a harmonic potential leads to a faster 
decay of the power law tails. \\
\hspace*{0.2cm}As mentioned earlier, the exponents of the long time tails that we obtain here for the 
position autocorrelation function are the same 
as one gets for a neutral particle in a viscous medium\cite{graberttwo}. However, the presence of the 
cyclotron frequency does modify the amplitudes of the 
power law tails of a charged particle 
in a magnetic field. Our predictions can be tested via cold atom-ion experiments\cite{Barkai,Katori,Sagi,Grimm,posvelprl,bhar2021} probing the position autocorrelation, the position-velocity 
correlation and the velocity autocorrelation functions of a charged particle in a magnetic field. 





 \bibliography{supurnareferences.bib}
  \end{document}